# Phase-controlled coherent photons for the quantum correlations in a delayed-choice quantum eraser scheme

Byoung S. Ham[1,2]
[1]School of Electrical Engineering and Computer Science, Gwangju Institute of Science and Technology, 123 Chumdangwagi-ro, Buk-gu, Gwangju 61005, South Korea
[2]Qu-Lidar, 123 Chumdangwagi-ro, Buk-gu, Gwangju 61005, South Korea
(October 20, 2023; bham@gist.ac.kr)

**Abstract**
The delayed-choice quantum eraser has been intensively studied for the wave-particle duality of a single photon in an interferometric system over the last decades. Coincidence measurements between quantum erasers have also been applied for the nonlocal quantum feature, satisfying the Bell inequality violation. However, those quantum features have not been clearly understood yet, resulting in the quantum mystery. Recently a coherence approach has been tried for the quantum eraser to unveil the quantum mystery. Here, a phase quantization of higher-order intensity products between coherently controlled quantum erasers is presented using a quarter wave plate-induced phase shift between orthogonal polarization bases of a single photon. Theoretical solutions of both photonic-de-Broglie-wave-like quantum features and nonlocal quantum correlations are presented for further discussions of quantum mechanics.

**Introduction**

Quantum mechanics has been developed based on the wave-particle duality [1] of a single particle over the last century, resulting in various quantum technologies in computing [2-4], communications [5-7], and sensing areas [8-10]. In a single photon's self-interference [11], quantum superposition between orthonormal bases of the single photon plays an essential role [12-15]. The wave-particle duality originated in quantum superposition must be exclusive in their natures according to the Copenhagen interpretation [16-18]. In that sense, the delayed-choice quantum eraser [12-15,19] can be understood as an ad-hoc quantum superposition of a measured photon through a dynamic window of a polarizer [19]. Due to the exclusive nature of the wave-particle duality, the violation of the cause-effect relation in the quantum eraser can also be understood for the superposition of many waves resulting in the bandwidth-limited effective coherence. Beyond the ensemble coherence, no quantum feature exists [19].

Quantum entanglement is between two or more individual photons (particles) whose fixed phase relation between paired photons does not violate quantum mechanics. A typical entangled photon pair is obtained by a spontaneous parametric down-conversion (SPDC) process [20] of the second-order nonlinear optics [21]. Because the wave mixing process of SPDC is coherent, a phase matching among the pump and two sibling photons is an inevitable birth condition [20,21,22,23]. In that sense, an assumption of a specific phase relation between the entangled photons is not absurd. Recently, such an understanding has been presented for the Hong-Ou-Mandel (HOM) effect [24,25] and Franson-type nonlocal correlation [26] using the wave nature. Experimental demonstrations have also been conducted in trapped ions for a $\pi/2$ phase difference [27]. A complete analytical solution of the phase relation between entangled photons shows the same $\pi/2$ for the HOM effect [25]. For the nonlocal correlation even between entangled photons, a coincidence measurement process for a particular product-basis selection is an essential condition [26]. Thus, the quantum feature should be understood differently from quantum entanglement itself.

Here, the fundamental physics of quantum features is presented for the PBW-like quantum features [28-30] and nonlocal correlations [31-34] using coherence manipulations of Poisson-distributed photons in a modified quantum eraser scheme. For the coherence manipulations, a quarter-wave plate (QWP) and a pair of acousto-optic modulators (AOMs) are used to control the phase of a single photon. For the PBW-like quantum feature, a QWP-caused phase shift between orthogonal polarization bases of a coherent photon is adopted to result in fringe multiplications for higher-order intensity products, showing the Heisenberg limit in resolution. For the coherently excited nonlocal quantum features between quantum erasers with and without QWP, a selective measurement technique is adopted for the intensity products using AOM-induced frequency-polarization correlations, resulting in the inseparable intensity product. Both features are known to be impossible by any classical means.

**Result**
*Phase control of coherent photons in a quantum eraser scheme*
Figure 1 shows the schematic of a modified delayed-choice quantum eraser using phase controls of Poisson-distributed photons with, firstly, QWP for the PBW-like quantum feature and, secondly, with a pair of AOMs for the nonlocal quantum feature. The photon number entering the noninterfering Mach-Zehnder interferometer (NMZI) is post-determined by coincidence detection [19]. The PBS of the NMZI provides distinguishable photon characteristics, resulting in no fringes in the NMZI output ports. For the polarization manipulation of the distinguishable photons, a polarizer (P) is added before each single-photon detector $D_j$ (j=1~4), resulting in the quantum eraser [19]. Thus, the particle nature of the photon retrospectively turns out to be the wave nature in the first-order intensity correlation [12-15]. If the physical distances between PBS and Ps are beyond the light cone, the quantum eraser in Fig. 1 satisfies the violation condition of the cause-effect relation [13]. This condition is also equivalently accepted for the violation of local realism in the second-order intensity correlation [34].

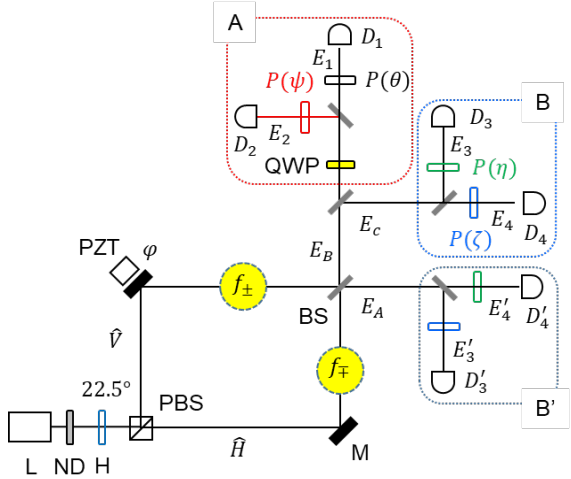

**Fig. 1. Schematic of coherence manipulations of coherent photons for a photonic de Broglie wave in a delayed-choice quantum eraser.** L: laser, ND: neutral density filter, H: half-wave plate, PBS: polarizing beam splitter, PZT: piezo-electric transducer, M: mirror, BS: nonpolarizing 50/50 beam splitter, QWP: quarter-wave plate, P: polarizer, D: single photon detector, H: horizon polarization, V: vertical polarization. $f_\pm$ and $f_\mp$ are generated by a pair of acousto-optic modulators. All rotation angles of Ps are independent of each other, satisfying local realism.

For the first task of the PBS-like quantum feature overcoming the diffraction limit in resolution, a quarter-wave plate (QWP) is inserted in one of the output ports of NMZI (see block A). As a result, an opposite fringe shift occurs between quantum erasers with and without QWP to the quantum erasers in block B or B' (see Fig. 2). Such a fringe shift is unprecedented in the first-order intensity correlation in both classical and quantum physics. This fringe shift is caused by the QWP-induced phase shift between orthogonal polarization bases of a single photon (see *Analysis*) [35]. At this point, first, the second-order intensity correlation between the output photons from either the same side (between blocks A and B) or opposite sides (between blocks A and B') is analyzed for the direct intensity products (see *Analysis*). As a result, a PBW-like quantum feature is obtained for fringe doublings proportional to the intensity-product order (N), satisfying the Heisenberg limit in resolution (see Fig. 3). Secondly, intensity products between two quantum erasers with and without QWP are post-controlled to discard (or choose) particular product bases. For this, a gated heterodyne detection technique [36] is adopted for the AOM-induced frequency-polarization correlation of paired photons at $f_\pm = f_0 \pm \delta f$, where the quantum state inside the NMZI is represented by $|\Psi\rangle = |f\rangle_\mp |H\rangle + e^{i\varphi}|f\rangle_\pm |V\rangle$.

The quantum eraser retrospectively reverses the predetermined particle nature of a photon into the wave nature via the action of the polarizer P [14,19]. Although the quantum eraser is known as one of the mysterious quantum phenomena [13-15] its perfect coherence understanding has been recently accomplished, where the violation of the cause-effect relation is due to the selective measurement of the orthogonal photons into a common polarization axis of the polarizer at the cost of 50% photon loss [19,37]. In the particle nature, this phenomenon is understood as an ad-hoc quantum superposition of decomposed orthogonal components of $\hat{H}$ and $\hat{V}$ whose common axis is along the polarizer, resulting in cancellation of the other decomposed components (see Section A of the Supplementary materials). The coherence solution of the quantum eraser for the first-order intensity correlation has no objection to the quantum approach due to the analytical equality to the classical one [18,38]. For the intensity product between output photons (quantum erasers) of NMZI, the phase quantization of the product basis is presented below to show the origin of the PBW-like quantum feature.

The QWP inserted output port of NMZI in Fig. 1 induces a phase retardation of $\pm \pi/2$ to the vertical polarization basis of a photon to the horizontal one, where the $\pm$ signs depend on the principle axis' rotation angles (0, $\pi/2$) of the QWP [35]. This polarization-dependent phase retardation should directly affect the quantum eraser because the role of the polarizer P is to project orthogonal polarization bases onto the common axis (see Eqs. (3)-(6)) [19]. In general, an interferometer is insensitive to the global phase of a photon due to Born's rule, where the measurement is the absolute square of the probability amplitude [12]. Interestingly, the phase control of the output photon by QWP directly results in the fringe shift of the quantum eraser (see Fig. 2). Such a fringe shift in the QWP-modified quantum eraser can be applied for phase quantization of the intensity products, resulting in the PBW-like fringe doubling feature (see Fig. 3). This phenomenon is unprecedented in coherence optics. In quantum mechanics, such a fringe shift of the second-order intensity product not for $\varphi$ but for θ (ψ, η, or ζ) has been the witness of the nonlocal quantum correlation (see Fig. 4) [39].

*Analysis 1: PBW-like quantum feature*
A coherence approach based on the wave nature of a photon is adopted to analyze Fig. 1 differently from the quantum approach based on quantum operators [40]. This means that the present *Analysis* is classical. For a normal quantum eraser scheme without QWP and AOMs, the amplitudes of output fields from the NMZI in Fig. 1 are represented by:

$$\boldsymbol{E}_A = \frac{iE_0}{2}\left(\hat{H} + \hat{V}e^{i\varphi}\right), \tag{1}$$

$$\boldsymbol{E}_B = \frac{E_0}{2}\left(\hat{H} - \hat{V}e^{i\varphi}\right), \tag{2}$$

where $E_0$ is the amplitude of a single photon from an attenuated laser L. $\hat{H}$ and $\hat{V}$ are unit vectors of horizontal and vertical polarization bases of a photon, respectively. Due to the orthogonal bases, Eqs. (1) and (2) result in no φ-dependent fringes, resulting in $\langle I_A \rangle = \langle I_A \rangle = I_0/2$ [19]. By the rotated polarizers whose rotation angles are from the horizontal axis, Eq. (2) is modified for the following quantum erasers:

$$\boldsymbol{E}_1 = \frac{E_0}{4}\left(\hat{H}\cos\theta - \hat{V}\sin\theta e^{i\varphi}\right)\hat{p}, \tag{3}$$

$$\boldsymbol{E}_2 = \frac{-iE_0}{4}\left(\hat{H}\cos\psi + \hat{V}\sin\psi e^{i\varphi}\right)\hat{p}, \tag{4}$$

$$\boldsymbol{E}_3 = \frac{-E_0}{4}\left(\hat{H}\cos\eta - \hat{V}\sin\eta e^{i\varphi}\right)\hat{p}, \tag{5}$$

$$\boldsymbol{E}_4 = \frac{-iE_0}{4}\left(\hat{H}\cos\zeta + \hat{V}\sin\zeta e^{i\varphi}\right)\hat{p}, \tag{6}$$

where $\hat{p}$ is the axis of the polarizers. In Eqs. (3)-(6), the insertion of meaningless $\hat{H}$ and $\hat{V}$ is just to indicate the photon's origin to understand the role of Ps, where only $\hat{H}$ is reversed by the BS, as shown in the mirror image. For the block B', $\mathbf{E}'_3 = \frac{E_0}{4}\left(\hat{H}\cos\eta - \hat{V}\sin\eta e^{i\varphi}\right)\hat{p}$ and $\mathbf{E}'_4 = \frac{iE_0}{4}\left(\hat{H}\cos\zeta + \hat{V}\sin\zeta e^{i\varphi}\right)\hat{p}$ are resulted. Due to the no action of the global phase on measurements by Born's rule, $\langle I'_3 \rangle = \langle I_3 \rangle$ and $\langle I'_4 \rangle = \langle I_4 \rangle$ are obtained. In other words, the quantum eraser schemes of blocks B and B' are identical for their fringes.

Thus, the corresponding mean intensities are described as follows for $\eta = \theta$ and $\zeta = \psi$:

$$\langle I_1 \rangle = \langle I_3 \rangle = \langle I'_3 \rangle = \frac{I_0}{16}(1 - \sin2\theta\cos\varphi), \tag{7}$$

$$\langle I_2 \rangle = \langle I_4 \rangle = \langle I'_4 \rangle = \frac{I_0}{16}(1 + \sin2\psi\cos\varphi). \tag{8}$$

The quantum mystery of the cause-effect relation of the quantum eraser can be found in the ad-hoc basis superposition of a single photon determined by $\hat{p}$ of the polarizer at the cost of 50% photon loss (see Section A of the Supplementary Materials). The global phase in Eqs. (3)-(6) does not affect the intensity fringe by Born's rule, as shown in Eqs. (7) and (8). For the balanced polarization control of the polarizer at $\theta = \psi = 45°$, $\langle I_1 \rangle = \langle I_3 \rangle = \frac{I_0}{16}(1 - \cos\varphi)$ and $\langle I_2 \rangle = \langle I_4 \rangle = \frac{I_0}{16}(1 + \cos\varphi)$ are resulted, as in the usual MZI, where fringes appear at $\varphi = 2n\pi$ (see Fig. 2). The fringe inversion between Eqs. (7) and (8), is due to the π-phase shift between $\hat{V}$ components. Recently a complete coherence solution of the quantum eraser has been proved for a single photon [19].

Now, a modified quantum eraser of Fig. 1 is analyzed with an inserted QWP in one output port of NMZI, resulting in an opposite fringe shift between quantum erasers in that port (see Fig. 2). This QWP-induced fringe shift is the origin of the phase quantization for the higher-order intensity products, resulting in the PBW-like quantum feature (see Fig. 3). In coherence optics, a QWP whose slow-axis is horizontal (0°) induces a phase gain of $\pi/2$ to the $\hat{V}$ component [35]. Thus, Eqs. (3) and (4) are written for the slow-axis horizontal (SA-H) QWP as follows:

$$\boldsymbol{E}_{1Q} = \frac{E_0}{4}\left(\hat{H}\cos\theta - i\hat{V}\sin\theta e^{i\varphi}\right)\hat{p}, \tag{9}$$

$$\boldsymbol{E}_{2Q} = \frac{-iE_0}{4}\left(\hat{H}\cos\psi + i\hat{V}\sin\psi e^{i\varphi}\right)\hat{p}. \tag{10}$$

The corresponding intensities are given by:

$$\langle I_{1Q} \rangle = \frac{I_0}{16}(1 + \sin2\theta\sin\varphi), \tag{11}$$

$$\langle I_{2Q} \rangle = \frac{I_0}{16}(1 - \sin2\psi\sin\varphi). \tag{12}$$

For $\theta = \psi = 45°$, Eqs. (11) and (12) show $\mp\pi/2$ phase-shifted fringes with respect to Eqs. (7) and (8), respectively (see the arrows in the lower left panel of Fig. 2). For the QWP whose slow-axis is vertical (90°), the fringes in Eqs. (11) and (12) are revered due to the sign reversal in $\hat{V}$ component. This fringe shift for the first-order intensity correlation is unprecedented in both classical and quantum physics.

For a generalized case with an arbitrary angle $\xi$ of the QWP at SA-H, the orthogonal polarization bases are represented by $\hat{H} \to \hat{H}e^{-i2\xi}$ and $\hat{V} \to \hat{V}e^{i2\xi}$ [35]. Then, Eqs. (9) and (10) are rewritten as:

$$\mathbf{E}_{1Q}^{\xi} = \frac{E_0}{4} e^{-i2\xi} \left( \hat{H} \cos\theta - i\hat{V} \sin\theta e^{i(\varphi+4\xi)} \right) \hat{p}, \tag{13}$$

$$\mathbf{E}_{2Q}^{\xi} = \frac{-iE_0}{4} e^{-i2\xi} \left( \hat{H} \cos\theta + i\hat{V} e^{i2\xi} \sin\theta e^{i(\varphi+4\xi)} \right) \hat{p}. \tag{14}$$

Thus, the corresponding intensities are obtained as:

$$\langle I_{1Q}^{\xi} \rangle = \frac{I_0}{16} (1 + \sin 2\theta \sin(\varphi + 4\xi)), \tag{15}$$

$$\langle I_{2Q}^{\xi} \rangle = \frac{I_0}{16} (1 - \sin 2\theta \sin(\varphi + 4\xi)). \tag{16}$$

For $\xi = 0°$, $\langle I_{1Q}^{\xi=0} \rangle = \frac{I_0}{16}(1 + \sin 2\theta \sin\varphi)$ and $\langle I_{2Q}^{\xi=0} \rangle = \frac{I_0}{16}(1 - \sin 2\psi \sin\varphi)$ confirm Eqs. (11) and (12).

Figure 2 shows the numerical calculations of Eqs. (15) and (16). As shown in the upper panels, the relations between Eqs. (15) and (16) are shown. In the lower-left panel, Eqs. (11) and (12) are confirmed for the fringe shifts of $\mp\pi/2$ with respect to the reference of Eqs. (7) and (8), as indicated by arrows. For $\xi = 45°$, $I_{1Q}^{\xi=45} = I_{2Q}^{\xi=0}$ and $I_{2Q}^{\xi=45} = I_{1Q}^{\xi=0}$ are resulted, as shown in the lower right panel. Thus, there is no difference between $\xi = 0°$ and $\xi = 90°$ for the QWP-modified quantum eraser.

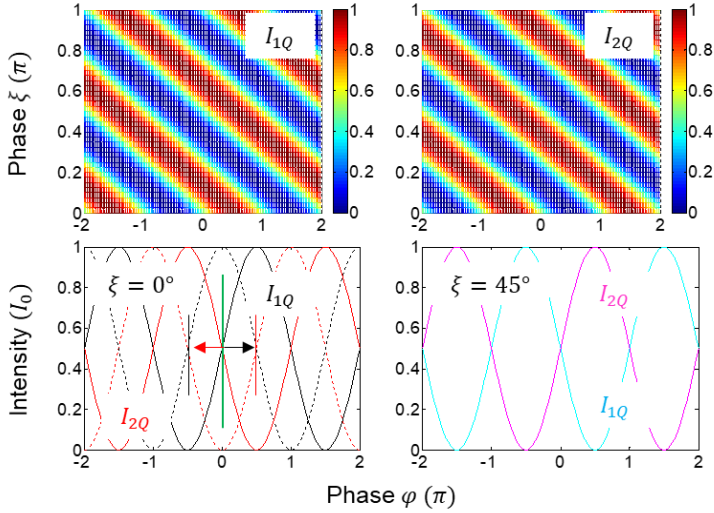

**Fig. 2. Numerical calculations of the QWP-induced fringe shift in the quantum eraser.** (lower left panel) black and red dotted curves are for Eqs. (7) and (8), respectively. $\theta = \psi = \eta = \zeta = \pi/4$.

The direct intensity product between Eqs. (15) and (16) for doubly-bunched coherent photons (N=2) is described as for $\xi = 0$ and $\theta = \psi = \pi/4$ (see the upper panels of Fig. 3):

$$\langle C_{1Q2Q}^{\xi=0}(0) \rangle = \frac{I_0^2}{64} \cos^2\varphi, \tag{17}$$

Here, a common factor $\sqrt{2}$ is multiplied to Eqs. (9) and (10) for the two-photon condition of the doubly bunched photon case for the coincidence detection. Thus, the number of fringes is doubled as the photon number is doubled in Eq. (17). This is due to the phase quantization of the intensity product basis, where the effect is the same as entangled photon-based PBW, even though its particle nature-based interpretation is different [41]. Without QWP, no PBW-like fringe doubling occurs, unless cross-intensity products are measured between NMZI output ports (see Section B of the Supplementary Materials). Thus, Eq. (17) shows the quantum feature for N=2.

The fourth-order intensity product between all quantum erasers of Eqs. (7), (8), and (17) for $\theta = \psi = \eta = \zeta = \pi/4$ is as follows:

$$\langle C^{(4)}(0)\rangle = \frac{I_0^4}{256}\sin^2\varphi\cos^2\varphi, \quad (18)$$

where a common factor 2 is multiplied to Eqs. (9) and (10) for the four-photon condition. Again, Eq. (18) shows the same fringe doubling of Eq. (17), resulting in the quadruple fringes compared to Fig. 2 (see lower left panel of Fig .3). Obviously, all equations from (1) to (18) are for classical physics. However, the fringe doubling effect in Eqs. (11), (17), and (18) for N=1, 2, 4 is unprecedented in classical (coherence) physics, whose results are equivalent to those of PBSs in quantum physics [28-30], satisfying the Heisenberg limit in resolution [42,43]. Thus, the classical physics of diffraction limit whose resolution is proportional to the square root of N is overcome by the action of QWP in the quantum eraser scheme (see Fig. 3(d)). The detailed analysis of the classical version limited by the standard quantum limit for Fig. 1 without QWP is shown in Section C of the Supplementary materials.

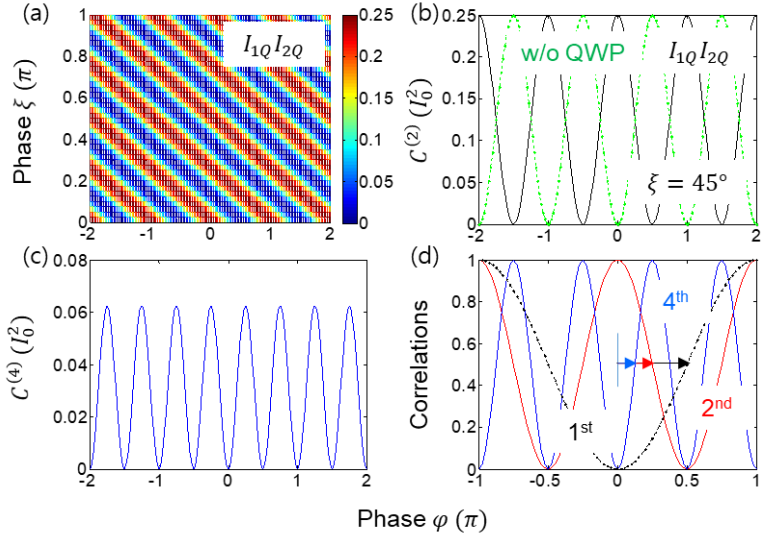

**Fig. 3. Numerical calculations for the phase quantization of the intensity product.** (middle panel) $C^{(2)} = C^{\xi}_{1Q2Q}$. (right panel) $C^{(4)}$ is the $C^{(2)}(Black) C^{(2)}(green)$ in the middle panel.

Figure 3 shows the numerical calculations of Eqs. (7), (8), (15), (16), and (18) for the PBW-like quantum feature. Figure 3(a) is for the second-order intensity correlation between Eqs. (15) and (16), showing the fringe doubling compared to Fig. 2. Figure 3(b) shows a $\pi/2$ fringe shift between the intensity products with (block A) and without (block B) QWP. Figure 3(c) is for Eq. (18) for the quadrupled fringes compared to Fig. 2. Finally, Fig. 3(d) is the PBW-like quantum feature whose resolution follows the Heisenberg's limit proportional to 1/N. For the classical resolution by the standard quantum limit ($\propto \frac{1}{\sqrt{N}}$), see Section C of the Supplementary Material.

*Analysis 2: Nonlocal correlation*

For the second-order intensity correlation between detectors $D_1$ (Alice) and $D_2$ (Bob), the coincidence measurements between Eqs. (9) and (10) are conducted by a gated heterodyne detection for the AOM-induced frequency-polarization correlated photon pairs, resulting in the selective choice of $\hat{H}\hat{V}$ product basis only:

$$\langle R_{Q1Q2}(0)\rangle = \frac{I_0^2}{64}\left(\hat{H}\cos\theta - i\hat{V}\sin\theta\right)\left(\hat{H}\cos\psi + i\hat{V}\sin\psi\right)(cc)$$

$$= \frac{I_0^2}{64}\hat{H}\hat{V}\sin^2(\theta - \psi), \quad (19)$$

where cc is a complex conjugate. For this gated heterodyne detection, the resolving time of a single photon detector must be much faster than the beating period to freeze the coincidence window [36]. Due to the action of AOMs resulting in the frequency-polarization correlation, i.e., $|\Psi\rangle = |f\rangle_\mp|H\rangle + e^{i\varphi}|f\rangle_\pm|V\rangle$, only $\hat{H}\hat{V}$ product is chosen for the DC-cut AC-pass filter, resulting in the inseparable intensity product in Eq. (19).

For this nonlocal quantum feature in Eq. (19), two independent polarizers $(\theta;\psi)$ are considered for $\xi = 0$ and $\varphi = \pi/2$ as a two-photon condition, satisfying local realism in both quantum erasers. Here, the polarizers act as a basis control as suggested by the Bell inequality test [44], and demonstrated for the Bell inequality violation [39]. The same result of Eq. (19) is also obtained for the usual MZI case of Eqs. (5) and (6) for the NMZI output port without QWP (see Section D of the Supplementary material). As analyzed for the Franson-type correlation [26], thus, the selective measurement process is the key to the nonlocal correlation. As shown in Section E of the Supplementary materials, however, no quantum feature exists between quantum erasers in blocks A and B, with and without QWP, i.e., between Eqs. (6) and (9) or Eqs. (4) and (10). Interestingly, thus, the phase relation between coincidently paired photons is quite important for the quantum feature (see lower left panel of Fig. 2) not only for the PBW-like feature but also for the nonlocal correlation in Eq. (19) (see Fig. 4). This also implies what the phase relation between higher-order entangled photons should be. In other words, Detectors 1 (2) and 3 (4) in Fig. 1 should be in the same party for the nonlocal correlation test, where the photons between them has a $\pi/2$ phase shift in fringes as shown in Fig. 2.

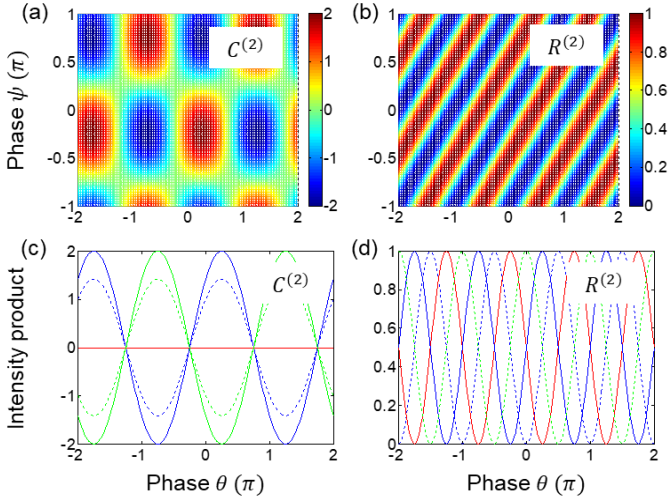

**Fig. 4. Numerical calculations of Eqs. (17) and (18).** $C^{(2)} = C_{1Q2Q}^{\xi=0}(0)$ in Eq. (17), $R^{(2)} = R_{Q1Q2}(0)$. (bottom panels) $\psi = -\frac{\pi}{4}$ (Blue); 0(Dotted); $\frac{\pi}{4}$ (Red); $\frac{\pi}{2}$ (Dotted green); $\frac{3\pi}{4}$ (Green). The polarizer's rotation angle θ (ψ) is for the path with (without) QWP in Fig. 1.

Figure 4 shows numerical calculations of both classical and quantum features derived in Eqs. (15), (16), and (19). Figure 4(a) is the direct intensity products between two output photons of Eqs. (15) and (16), whereas Fig. 4(b) is the result via a gated heterodyne detection for Eq. (19). Figures 4(c) and (d) are the corresponding details of Figs. 4(a) and (b), respectively. The fringe shift of the product basis in Fig. 4(d) is the witness of the Bell inequality violation [39]. Due to the selective choice of the product bases in Eq. (19), the cost to pay for the classical feature of Figs. 4(a) and (c) is 50% measurement event loss in the quantum feature in Figs. 4(b) and (d). In other words, the quantum features in Figs. 2, 3, and 4 are the quantum illusion caused by selective measurements of coherence tensor products at 50%. Even for the entangled photons, no nonlocal quantum correlation exists if no selective measurement is conducted by the polarizers [31-34,37] or coincidence [26,45].

**Discussion**

*Phase quantization of the intensity product vs. PBWs*

Using higher-order entangled photon (atom) pairs such as an N00N state, PBWs have been intensively studied over the last several decades [28-30,42,43]. In a quantum approach based on the particle nature of a photon, the N-proportional fringe multiplication in PBWs is understood due to the N-times repeated quantum operators for the same MZI path, resulting in the product-basis quantization at $\pi/N$ in $e^{iN\varphi}$ [42,43]. Here, N=1 is for the classical case. In Fig. 1, the same phase-basis quantization was derived from a coherence approach by fringe shifting of the quantum eraser for N=1 using a QWP, where the QWP-induced phase shift is between orthogonal polarization bases of a single photon. Obviously, such QWP-induced phase shift cannot contribute to the fringe shift without the polarizer in the quantum eraser due simply to the Born's rule. The polarizer in the Bell inequality test is basically for the quantum eraser because the entangled photons have an inherent $\pi/2$ phase shift between them [24,25,37], which is equivalent to the BS output photons [46]. Thus, the coherence solutions of the PBW-like quantum feature shed light on further discussions of the phase relation between higher-order entangled photons.

*Violation of the cause-effect relation*

As long as the cause-effect relation refers to information about a signal, monochromatic waves have no meaning as discussed in the superluminal propagation in the early 2000s [47]. Thus, a spectral bandwidth of photon pairs is a prerequisite to discuss the cause-effect relation. In that sense, the coherence approach must be extended for an AOM scanning mode to comply with the information concept. Then, the derived coherence solutions are simply for the linear superposition of the measurement events for the bandwidth-distributed photon pairs. Regarding the light cone of the cause-effect relation bounded by the measurement-time window of a coincidence counting module, the ensemble coherence of photon pairs limits the coincidence-timing range between them. To demonstrate the cause-effect relation, thus, narrow band-pass filtered SPDC-generated entangled photon pairs are necessary if the traveling time between the PBS (source) and polarizers in Fig. 1 should be longer than the ensemble coherence time. Even in this case, however, the violation of the cause-effect relation may have no practical meaning due to the selective choice of paired photons at the cost of 50% loss via the ad-hoc quantum superposition assigned by the polarizer's axis. The ad-hoc quantum superposition makes the system coherent, where the measured ones are not the original photons fixed at $\hat{H}$ and $\hat{V}$ polarization bases. As demonstrated by the quantum eraser, each photon's orthogonal polarization bases are redefined by the rotated polarizer in the name of polarization projection. This is the meaning of the ad-hoc quantum superposition. The violation of the cause-effect relation will be discussed further.

**Conclusion**

Using a phase control of attenuated photons in NMZI modified by QWP and polarizers, a PBW-like quantum feature was analytically demonstrated for the fringe doubling-caused N-proportional resolution enhancement via quantization of the intensity-product bases, where the QWP-induced a $\pi/2$ phase shift between orthogonal polarization bases of a single photon. Unlike usual coherence optics, this phase shift was post-picked up by the polarizer, resulting in the $\pm\pi/2$ fringe shifts with respect to non-QWP-based quantum erasers. Thus, the second (fourth)-order intensity correlation between them resulted in the fringe doubling (quadrupling) effect. This relates to the PBWs at N=2 (4). Thus, the analytically derived coherence solutions of the phase-controlled quantum erasers for the higher-order intensity correlations satisfied the Heisenberg limit in resolution, overcoming the classical diffraction limit governed by the standard quantum limit. For the nonlocal quantum correlation between the phase-controlled quantum erasers, a gated heterodyne detection technique was adopted for the product-basis selection in coincidence measurements, where the AOM pair-induced frequency-polarization photon pairs played the key role in the selective measurements of orthogonally polarized product bases. Related numerical calculations of both PBW-like quantum features and nonlocal correlations were perfectly consistent with the observed quantum features using entangled photons. Thus, the present study sheds light on a better understanding of quantum mechanics, where the phase relation between entangled photons is necessary to interpret the results to unveil the quantum mystery. Unlike nonlocal correlation, the PBW-like quantum feature does not require the violation of local realism. A generalized phase relation between paired photons is discussed elsewhere for a universal coherence solution of higher-order PBW-like quantum features applicable to quantum technologies including quantum sensing.

**Methods**

The NMZI in Fig. 1 comprises a polarizing beam splitter (PBS) and a 50/50 nonpolarizing beam splitter (BS). The coherent photon is randomly generated from the attenuated laser L by Poisson statistics, whose mean photon number is set to be $\langle n \rangle < 1$ to satisfy independent and incoherent photon statistics. For the higher-order intensity correlations between quantum erasers, bunched photons are randomly and statistically post-selected by a single-photon counting module [19]. The generation ratio of (N+1)-bunched photons to N-bunched photons is determined by Poisson statistics at a few percent. For the particular order of intensity correlation, the coincidence measurements through the single-photon counting module play an essential role [19]. For random polarizations of an input photon, a 22.5°-rotated half-wave plate (HWP) is added before NMZI, where the laser L is vertically polarized. For the AOM-induced frequency-polarization correlation between path-paired photons, an oppositely diffracted photon pair is chosen for a common rf frequency. The polarizer's angle is from the horizontal axis toward a counterclockwise direction [19]. For gated heterodyne detection, a DC-cut AC-pass filter is used to block the same polarization product bases of paired photons. Due to the random chance of transmission and reflection on either PBS or BS, all bunched photon cases in each BS are discarded for measurements of the second-order intensity correlation by the DC-cut AC-pass filter, resulting in the orthogonally polarized product-basis selection. Here, the terminology of 'nonlocal' is provided by the quantum eraser scheme, whose physical distance between the PBS and each polarizer is beyond the light cone, satisfying the violation condition of the cause-effect relation determined by the relativity theory.

**Acknowledgment**

The author thanks Prof. J. Stöhr at Stanford university for a helpful discussion.

**Funding:** This research was supported by the MSIT (Ministry of Science and ICT), Korea, under the ITRC (Information Technology Research Center) support program (IITP 2023-2021-0-01810) supervised by the IITP (Institute for Information & Communications Technology Planning & Evaluation). BSH also acknowledges that this work was also supported by GIST GRI-2023.

**Author contribution:** BSH solely wrote the paper.

**Competing Interests:** The author declares no competing interest.

**Data Availability Statement:** All data generated or analyzed during this study are included in this published article.